\documentclass[twocolumn,aps,prl,superscriptaddress,amssymb,showpacs]{revtex4}
\usepackage[dvips,final]{graphicx}
\begin{document}
\title{Experimental Evidence for Two-Dimensional Magnetic Order in Proton Bombarded Graphite}
\author{J. Barzola-Quiquia}
\author{P. Esquinazi}\email{esquin@physik.uni-leipzig.de}
\author{M. Rothermel}
\author{D. Spemann}
\author{T. Butz}
\affiliation{Institut f\"{u}r Experimentelle Physik II,
Universit\"{a}t Leipzig, Linn\'{e}stra{\ss}e 5, D-04103 Leipzig,
Germany} \author{N. Garc\'ia}
 \affiliation{Laboratorio de F\'isica de Sistemas Peque\~nos y Nanotecnolog\'ia,
 Consejo Superior de Investigaciones Cient\'ificas, E-28006 Madrid, Spain} 
\begin{abstract}
We have prepared magnetic graphite samples bombarded by protons at
low temperatures and low fluences to attenuate the large thermal
annealing produced during irradiation. An overall optimization of
sample handling allowed us to find Curie temperatures $ T_c \gtrsim
350$~K at the used fluences. The magnetization versus temperature
shows unequivocally a linear dependence, which can be interpreted as
due to excitations of spin waves in a two dimensional Heisenberg
model with a weak uniaxial anisotropy.
\end{abstract}
 \maketitle

Recent advances to develop nanographitic systems have led to a
renewed interest on their electrical properties worldwide\cite{1}.
A single layer of graphite, the two-dimensional (2D) graphene,
appears to have quantum properties at room temperature\cite{2} as
well as rectifying electronic properties\cite{3,4}. On the other
hand, some of those properties were already observed in highly
oriented pyrolytic graphite (HOPG) of low mosaicity, as the
quantum Hall effect\cite{5} and de Haas - van Halphen quantum
oscillations even at room temperature\cite{6}.
 The two-dimensional properties of the graphene planes in graphite open
up the possibility of using nanometer to micron size regions of
graphite in new integrated devices with spintronic properties either
through the use of ferromagnetic electrodes, e.g. spin-valves, and/or
making graphite itself magnetic. In fact this has been a topic of
study in the last years and reports exist showing magnetic hysteresis
in blank graphite\cite{10} but especially in proton bombarded
graphite\cite{11}. Severe limitations in the sensitivity and
reproducibility of standard magnetometers added to annealing effects
during bombardment, hindered the identification of a critical
temperature $T_c$ as well as the characteristics and dimensionality
of the ferromagnetic signals. The aim of this work is to show that
specially prepared highly oriented pyrolytic graphite (HOPG) samples
show ferromagnetic order with $T_c \gtrsim 350$~K and the
magnetization temperature dependence is in good agreement with  a 2D
anisotropic Heisenberg model (2DHM) and the presence of spin waves
excitations \cite{12,13,14}.

For the experiments we used four pieces of  a HOPG sample grade ZYA,
samples 1 to 4 (mass: 12.8, 12.5, 10.1, and 6~mg respectively)
irradiated by a 2.25 MeV proton micro-beam (sample 4: 2.0 MeV, 0.8~mm
broad beam) perpendicular to the graphite planes. With the micro-beam
we produced several thousands of spots of $\sim 2~\mu$m diameter each
and separated by $5~\mu$m (sample 1) or $10~\mu$m (samples 2 and 3)
distance, similarly to the procedure used in Ref.~\onlinecite{15}.
Samples 1 and 2 were irradiated at 110~K whereas samples 3 and 4 at
room temperature. Further irradiation parameters for sample 1 (2,3,4)
were: 51375 (25600,25600,6) spots, fluence: 0.124
(0.08,0.13,0.3)~nC/$\mu$m$^2$, total irradiated charge 46.9
(44.8,37.4,900)~$\mu$C, and 1~nA proton current (100~nA for sample
4).  The pieces we have irradiated showed an iron concentration (the
only detected magnetic impurity) within the first $35~\mu$m  of $\sim
(0.4\pm 0.04)~\mu$g/g ($< 0.1~$ppm).

Previous experiments \cite{11} showed ferromagnetic magnetic moments
at saturation $m_{\rm sat}\sim 1~\mu$emu and therefore put severe
constrains to experimentalists, not only regarding the sensitivity of
the used magnetometer but also its reproducibility after sample
handling. In this work two main experimental improvements have been
achieved. Firstly, we enhanced the ferromagnetic part produced by
irradiation reducing  annealing effects. In samples 1 and 2 the
micrometer spots were produced  at a nominal temperature of 110~K
during irradiation (18~hours). For comparison and to reduce further
annealing effects sample~4 was irradiated with a broad beam and low
fluence. Second, we have designed a sample holder that allows us to
measure the magnetic moment of the sample in the SQUID and to fix it
inside the irradiation chamber without any changes. We investigated
the reproducibility of the magnetic measurements and checked that the
sample holder handling (with sample) \cite{16}, i.e. inserting it and
taking it out of the irradiation and SQUID chambers \cite{squid},
does not produce systematic changes of the magnetic signal. Our
arrangement provides a reproducibility of $\sim 10^{-7}$~emu in the
measured field range and allows the subtraction of the virgin data
from those after irradiation point by point, increasing substantially
the sensitivity of the magnetic measurements to $\sim 2 \times
10^{-8}$~emu.

Figure 1 shows the hysteresis loops of the magnetic
\begin{figure}
\begin{center}
\includegraphics[width=80mm]{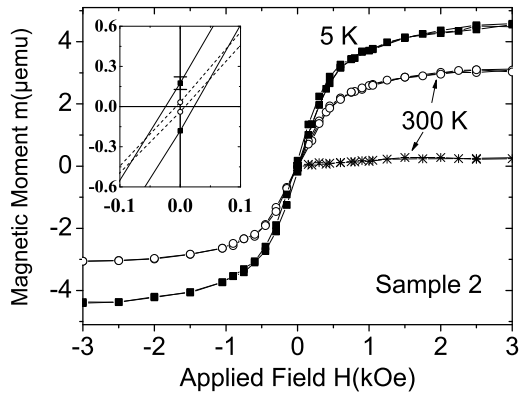}
\caption[]{Magnetic moment as a function of applied field for the
irradiated sample~2 at 300~K (o) and 5~K $(\blacksquare)$ obtained
after subtracting the data of the non irradiated sample. The
points ($\star$) are obtained for the same sample at 300~K after
taken out the first $\sim 5$ micrometers from the irradiated
surface side. The inset blows out the data at low fields to show
the finite hysteresis and the clear temperature dependence of the
coercive field and remanent magnetic moment.} \label{fig1}
\end{center}
\end{figure}
 moment $m$ of
sample 2 at two temperatures. These loops are obtained directly from
the difference of the measurement ``after" minus ``before"
irradiation. The loop at 5~K as well as the measured temperature
dependence at constant field indicate a paramagnetic contribution
$m_p = 0.575~H / T [\mu$emu~K/kOe] for this sample, i.e. less than
10\% of the ferromagnetic signal at 3~kOe. At 300~K, however, $m_p$
is negligible. These loops, their temperature dependence as well as
the finite hysteresis, see inset in Fig. 1,  indicate the existence
of magnetic order with a Curie temperature higher than room
temperature.

Sample~3, which was irradiated with similar number of spots, fluence
and total charge but at room temperature, shows a
 a ferromagnetic signal at saturation
  $\sim 5$ smaller than that  obtained
for samples 1 or 2, in agreement with previous work\cite{11}.
These results indicate the reliability and sensitivity of the used
procedure as well as the absence of obvious artifacts in the
measurements.

After peeling out the first micrometers from the irradiated surface
of sample~2 the ferromagnetic contribution decreased by one order of
magnitude, see Fig.~1. We can answer now the question whether the Fe
concentration in the sample and due to some hypothetical annealing by
the protons could be responsible for the observed ferromagnetic
signal. In the first micrometer and taking an irradiated area
$\lesssim 0.026~$cm$^2$, the magnetization at room temperature is
then $\gtrsim 0.5~$emu/g. In this region we estimate that the mass of
the ferromagnetic carbon material is $< 6~\mu$g. Were the measured Fe
concentration ferromagnetic at 300~K then it would contribute with a
magnetic moment $\lesssim 0.6 \times 10^{-10}$~emu, i.e. 50.000 times
smaller than the measured one. With the mass of the ferromagnetic
part of the irradiated HOPG sample we estimate a magnetic moment per
carbon atom $m_C \gtrsim 0.001~\mu_B$, in very good agreement with
XMCD results \cite{18}.
\begin{figure}
\begin{center}
\includegraphics[width=80mm]{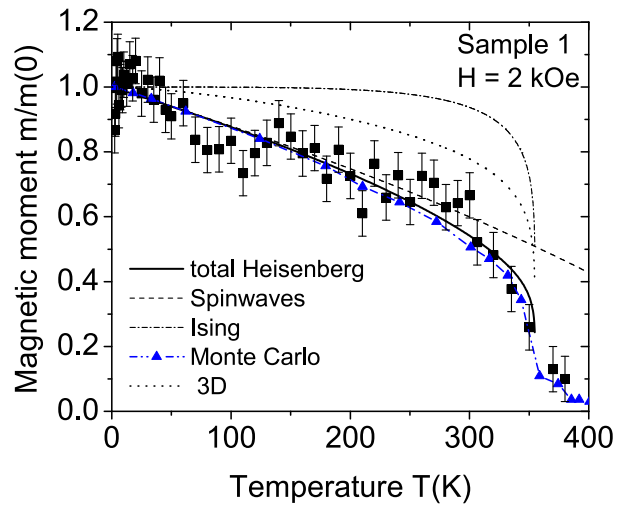}
\caption[]{Normalized magnetic moment $(m(0) = 2.60~\mu$emu) obtained
for irradiated sample~1 at  2~kOe. The data points are obtained after
subtracting the data from the sample before irradiation and a
paramagnetic (Curie) contribution $m_p(T) = 4.9 / T [\mu$emu K]. The
errors bars indicate typical errors due to the subtraction of the
data from the virgin sample. The chosen parameters for the
theoretical curves are $T_c = 360~$K, $T_c^{sw} = 850~$K ($\Delta =
0.001$). The continuous line is obtained from (\ref{mzt}). The dotted
line is the 3D Bloch $T^{3/2}$ model with spin waves
\protect\cite{kit7}. The dashed-dotted line with close triangles
shows the results of a Montecarlo simulation with anisotropy (square
lattice of $200 \times 200$~points).} \label{fig2}
\end{center}
\end{figure}

Figures 2 and 3 show the temperature dependence of the ferromagnetic
moment for samples 1 and 4, respectively. Because the paramagnetic
signal contributes significantly only at $T \lesssim 25~$K, we have
subtracted it in both figures in order to show only the ferromagnetic
part. Up to the highest measured temperature of 380~K this magnetic
moment behaves reversible. Furthermore, no changes in $m$ within
experimental error were observed after leaving the samples several
months at room temperature.

One of the interesting and indicative results shown in Figs.~2 and 3
is the unequivocal linear dependence. This is an indication of 2D
magnetism and the slope can be interpreted as due to the excitation
of 2D spin waves that reduce the magnetization linearly with T
\cite{12,13,14}. We are not aware of any model hamiltonian producing
such a linear behavior in $m(T)$. Therefore, to analyze the measured
temperature dependence we discuss the 2DHM with anisotropy that
provides a linear dependence with T. The discrete Hamiltonian
describing the 2DHM reads $ H = - J\sum_{ij}
[S_{iz}S_{jz}+(1-\Delta)(S_{ix}S_{jx} +S_{iy}S_{jy})]$,
 where $S_i=(S_{ix},S_{iy},S_{iz})$ represents a
unit vector in the direction of the classical magnetic moment
placed at the site $i$ of a 2D lattice. The sum $(i,j)$ is
performed over all nearest neighbor pairs, and $J$ is the exchange
coupling. The parameter $\Delta$ represents the uniaxial
anisotropy in the $z$-direction. The case $\Delta= 0$ is the
isotropic 2DHM and is known to have $T_c=0$. However, just a small
anisotropy raises $T_c$ considerably because $T_c\sim
-1/\ln\Delta$ for $\Delta \rightarrow 0$.

It can be shown \cite{12,13,14} that the normalized spin-waves
magnetization in the anisotropic axis behaves as $ M_z^{sw}=1-
T/T_c^{sw} - 2T^2/(T^\star T_c^{sw}) - (2/3)(T/T_c^{sw})^3$ at low
temperatures, where $T^\star = 4 J$. This result is obtained using
perturbation theory techniques\cite{19,20} up to third order in spin
waves. The parameter
 $T_c^{sw}$
is the spin wave critical temperature due to low-energy spin wave
excitations; it is given by $k_BT_c^{sw}=2\pi J/K(1-\Delta)$,
where $K(x)$ is the elliptic function. Near the critical
temperature $T_c$ the physics can be better described by a 2D
Ising model that should provide a good description of the spin
flip excitations. Then $T_c$ is given by
$T_c(\hat{J})=1.52\hat{J}$\cite{19}, where $\hat{J}$ is the
renormalized exchange due to the spin waves excitations according
to the expression $\hat{J}(T)=J[1-2T/T_c^{sw}]$. The values of
$M_z$ at $T<T_c$ can be expressed as:
\begin{equation}
M_z(T) \approx M_z^{sw}(T,J)M_z^I[T,\hat{J}(T)]\,. \label{mzt}
\end{equation}
 The first factor in the rhs of (\ref{mzt}) is the magnetization due
to spin waves and the second one is the magnetization due to an Ising
model with the exchange renormalized by the spin waves. We have
checked this theoretical result against Montecarlo calculations with
$\Delta=0.001$ and the agreement is excellent, especially at low
anisotropies\cite{14} as it is shown in Figs.~2 and 3.  In Fig.~2 we
have plotted also the normalized spin waves contribution
$M_z^{sw}/M_z^{sw}(0)$ up to third order. The Heisenberg result
approximated by (\ref{mzt}) and the Montecarlo calculation agree and
both fit the experimental data with the parameters $T_c^{sw} =850~$K,
$\hat{J}(T_c=360$K$)=237$~K, indicating an anisotropy $\Delta \simeq
0.001$. Sample~2 shows a similar behavior and its data can be fitted
with $T_c^{sw} \simeq 1000~$K, $\hat{J}(T_c \simeq 310$K$) = 202$~K.
The data for sample 4 shown in Fig.~3 show also a linear behavior.
Extrapolating the SW contribution to $m(T^\star) \simeq 0$ we
conclude that $T_c < T^\star \simeq 640~$K. Then using (1) we
estimate $T_c \gtrsim 450~$K with $\Delta \lesssim 10^{-4}$, see
Fig.~3. These results already show that $T_c$ increases with fluence,
provided that one can reduce simultaneously the annealing effects
produced during irradiation. For comparison we also have plotted in
Figs. 2 and 3 the Ising model result that has no spin waves and the
3D Bloch $T^{3/2}$ law that includes spin waves \cite{kit7}. The
comparison indicates clearly that spin waves in 2D dominate the
magnetization up to $\gtrsim 300$~K and that the usual 3D model does
not fit the data.

\begin{figure}
\begin{center}
\includegraphics[width=80mm]{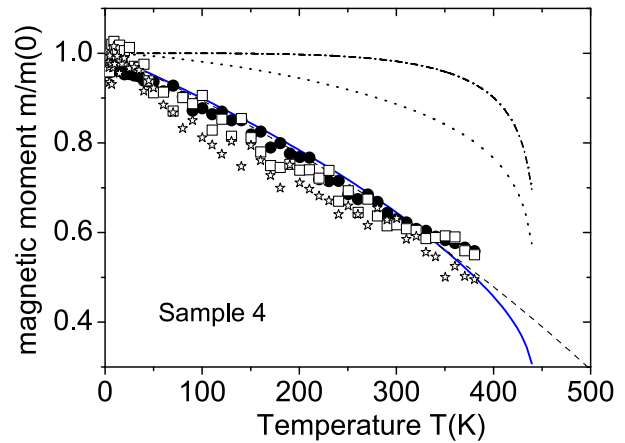}
\caption[]{Normalized magnetic moment ($m(0) = 4.9~\mu$emu at 10~kOe)
obtained for sample~4 at (10,3,1)~kOe $(\bullet,\square,\star)$ after
subtracting the data from the sample before irradiation and a
paramagnetic (Curie) contribution $m_p(T) = 1.18 H / T [\mu$emu
K/kOe]. Different theoretical curves are the same as in Fig.~2 but
with parameters: $T_c = 450~$K and $T_c^{sw} = 1050~$K.} \label{fig3}
\end{center}
\end{figure}

There is no doubt that defects in the graphite structure are one of
the possible origins for localized magnetic moments. The
ferromagnetism triggered by the bombardment should be correlated to
the produced defects located at approximately the first micrometer
from the sample surface. To discuss a mechanism responsible for the
coupling between the magnetic moments, we need first to estimate the
density of defects. For sample 1 we have 0.9~nC total irradiated
charge per spot in an area of ($\sim \pi 0.6^2)\mu$m$^2$. Using
SRIM2003 Monte Carlo simulations with full damage cascades and 35~eV
displacement energy we obtain a vacancy density of $\sim 5 \times
10^{20}~$~cm$^{-3}$ at the surface, which means a distance between
vacancies $l \sim 1.3~$nm$\sim 9 a$, where $a = 0.14~$nm. This
distance is much smaller than the inverse of the Fermi wave vector
$1/k_F \sim 30~$nm for a Fermi energy of 20~meV or calculated using
the 2D carrier density\cite{5}.

Regarding the coupling needed to have room temperature magnetic
ordering there is in first place the direct coupling for nearly
localized spins at the defects, which should be in the range of $\sim
300~$K. Recently the RKKY coupling between large defects in graphene
has been studied for Fermi energy tending to zero\cite{21}. This
coupling might be always ferromagnetic because $k_F r \ll 1$ for $r
\sim l$. However, estimations of the Curie temperature for this
coupling within our defect densities provide values of the order of
20~K. What appears important is a super-exchange mediated by the two
different sites in the graphite lattice\cite{22,23} or between
magnetic moments from defects and from hydrogen atoms, which may
effectively increase the magnetic moment density on a graphene
lattice.

We note that large concentration of hydrogen is found in the first
micrometer thick region at the surface of graphite samples\cite{24}.
Therefore we should take into account the possible influence of
hydrogen in triggering localized as well as non-localized magnetic
moments in the graphite layers\cite{25,22}. Irradiation may
contribute as defect generation as well as dissociating the existing
molecular hydrogen enabling its diffusion and bonding in defective
parts of the lattice structure. All these moments will tend to be
ferromagnetically coupled enhancing the Curie temperature by the RKKY
coupling.

 Within this picture it becomes clear that the enhancement of
 the defect density, which occurs at larger depths from the
 surface in the inner part of the irradiation path up to
full amorphization at a depth $\sim 35\ldots 40~\mu$m, perturbs too
much
 the graphene lattice destroying in this way the necessary
 band structure and carrier density. This may
 explain the experimental observation of a rather
 well-defined critical temperature (and not a distribution) and
 also the difficulty one has to reach much higher ferromagnetic
 magnetization values increasing the proton fluences clearly
 above the values used here. If an electron-mediated
 coupling between defects plays a role, we expect that for
 an adequate defect density it should be possible to
 influence the magnetic order shifting the Fermi energy
 by applying an appropriate bias voltage.

The results of samples~1 and 2 provide clear evidence for the good
reproducibility of our approach: although the spot density, beam
diameter as well as total charges were different, the produced defect
densities in the irradiated paths were similar for both samples and
therefore we expect to obtain similar critical temperatures as the
measurements showed. Changing the defect density as well as their
distribution in the lattice one may tune the ferromagnetic transition
temperature as well as the magnitude of the magnetization produced by
irradiation, as the data for sample~4 clearly indicate. As a rule of
tumb robust ferromagnetism with $T_c > 300~$K by proton irradiation
in graphite can be reached with fluences of the order of
0.1~nC/$\mu$m$^2$.

In conclusion our work shows that  irradiation of micrometer spots in
graphite at low temperatures as well as broad irradiation, both at
very low fluences, increases significantly the magnitude of the
magnetic order with Curie temperatures $T_c \gtrsim 300$~K. The use
of especial sample holders made possible to reduce sample handling
between irradiation chambers and SQUID measurements to a minimum,
ruling out simple introduction of impurities or the influence of
operative artifacts. This approach increased substantially the
sensitivity and reproducibility of the magnetization measurements
allowing us to obtain directly the effects produced by irradiation
within an error of $\sim 10^{-7}$~emu. The experimental localization
of the ferromagnetic irradiated part of the sample indicates that the
graphite structure is important and that at the used proton energies
low fluences are preferential to trigger a robust ferromagnetic
order. We showed that the magnetization of the magnetically ordered
contribution decreases linearly at  $T < T_c$, a behavior that can be
assigned to the signature of low energy spin waves excitations well
described by an uniaxial two dimensional anisotropic Heisenberg
model.

We gratefully acknowledge discussions with M. A. Vozmediano and L.
Pisani. This work was done in the framework of the EU project
``Ferrocarbon" and partially supported by the DFG under ES 86/11-1.


\begin{thebibliography}{10}
\bibitem{squid} Our
superconducting quantum interferometer device (SQUID) measurements
were done using a reciprocal sample option (RSO), which increases the
sensitivity by a factor of ten. The magnetic field was applied
parallel to the graphene planes.
\end{thebibliography}
\end{document}